\documentclass[letterpaper, twoside, 10 pt, conference]{ieeeconf}  
\usepackage{fancyhdr}       
\usepackage{amssymb}
\usepackage{graphicx}
\usepackage{algorithm}
\usepackage{algcompatible}
\usepackage{amsmath}

\IEEEoverridecommandlockouts                              
\def\Prob{{\rm{Prob}}}
\def\ecp{{\rm{ecp}}}
\def\becp{\overline{{\rm{ecp}}}}

\def\ep{{\rm{ep}}}
\def\by{\bar{y}}

\def\qp{\tau} 

\def\cD{{\mathcal{D}}}
\def\cV{{\cal{V}}}

\def\cX{{\mathcal{X}}}
\def\cY{{\mathcal{Y}}}

\def\by{{\bar{y}}}

\def\tx{{\tilde{x}}}
\def\ty{{\tilde{y}}}

\newcommand {\bsis} {\left\{ \begin{array} }
\newcommand {\esis} {\end{array}\right.}
\def\Sum#1#2{\sum\limits_{#1}^{#2}}
\def\R{ {\rm \,I\!R} }    
\def\fracg#1#2{{\displaystyle{\frac{#1}{#2}}}}  
\def\set#1#2{\{ \; #1 \;:\;#2\;\}} 

\newcommand {\bmat} {\left[\begin{array} }

\newcommand {\emat} {\end{array}\right]}

\newcommand{\bv}{\bmat{c}}
\newcommand{\ev}{\emat}

\newcommand {\T}{^{\top}} 
\newcommand{\blista}{\renewcommand{\labelenumi}{(\roman{enumi})} 
\begin{enumerate}}
\newcommand{\elista}{\end{enumerate} \renewcommand{\labelenumi}{\arabic{enumi}.}}

\newtheorem {definition}{Definition}

\newtheorem {property}{Property}
\newtheorem {remark}{Remark}

\def\Id{\rm{I}}

\title{\LARGE \bf Probabilistic interval predictor based on dissimilarity functions}

\author{A. D. Carnerero, D. R. Ramirez and T. Alamo
\thanks{This research has been funded   by  Ministerio  de  Econom\'ia  y  Competitividad  of Spain under project DPI2016-76493-C3-1-R and by Ministerio de Ciencia e Innovación of Spain under project PID2019-106212RB-C41.}
\thanks{Departamento de Ingenier\'ia de Sistemas y Autom\'atica. Universidad de Sevilla. Spain {\tt\small acarnerero,danirr,talamo@us.es}}%
\thanks{\copyright{}2021 IEEE. Personal use of this material is permitted.  Permission from IEEE must be obtained for all other uses, in any current or future media, including reprinting/republishing this material for advertising or promotional purposes, creating new collective works, for resale or redistribution to servers or lists, or reuse of any copyrighted component of this work in other works.~\hfill}%
}

\begin{document}

\maketitle
\pagestyle{empty}
\thispagestyle{fancy}

\begin{abstract}
This work presents a new methodology to obtain probabilistic interval predictions of a dynamical system. The proposed strategy uses stored past system measurements to estimate the future evolution of the system. The method relies on the use of dissimilarity functions to estimate the conditional probability density function of the outputs. A family of empirical probability density functions, parameterized by means of two scalars, is introduced. It is shown that the proposed family encompasses the multivariable normal probability density function as a particular case. We show that the presented approach constitutes a generalization of classical estimation methods. A validation scheme is used to tune the two parameters on which the methodology relies. In order to prove the effectiveness of the presented methodology, some numerical examples and comparisons are provided.
\end{abstract}



\section{Introduction} \label{section1}

Consider a discrete nonlinear system
\begin{equation}
y_{k} = f_0(x_k,w_k),\label{sistema}
\end{equation}
where $f_0(\cdot,\cdot)$ is not known, $k$ is the discrete time instant, $y_k\in \cY \subseteq \R$ is the output of the
system, $w_k$ accounts for parametric uncertainty, noise, disturbances, etc. Also, vector $x_k \in X \subseteq \R^{n_x}$ represents the past inputs and
outputs of the system, i.e., $x_k=
[y_{k-1},y_{k-2},...,y_{k-n_y},u_{k},u_{k-1},...,u_{k-n_u}]$ and $n_x = n_y+n_u+1$. Note that nonlinear terms of past system inputs-outputs could be incorporated into vector $x_k$.

In this paper we focus on interval predictions. That is, given the regressor $x_k$, the objective is to compute an interval $I(x_k)=[y_k^-,y_k^+]$ such that we maximize the probability that $y_k$ belongs to $I(x_k)$ while minimizing the interval width $(y_k^+-y_k^-)$. These two conflicting objectives can be reconciled if one minimizes the interval width with the constraint that $I(x_k)$ contains $y_k$ with a pre-specified probability.

Interval predictions play a relevant role in the control of uncertain systems. Zonotopes and DC Programming are used to obtain interval state estimators in \cite{AlamoAUT05} and \cite{AlaBravRedCama08} respectively. Interval observers for linear time-varying systems have been proposed in \cite{Thabet20142677} and \cite{Chebotarev201582}. Fault detection methods based on zonotopic bounds can be found in \cite{Raka2013119}. In  \cite{Xu2014947},  set theoretic approaches are also used in the context of fault detection. Set membership methods  \cite{milanese2004set,milanese2011unified} can also be used to obtain interval predictions. A mixed Bayesian/set-membership approach is proposed in \cite{FernandezCanti201559}. 

There exists different methods in the literature that address the problem of obtaining interval predictions for system (\ref{sistema}). For example, if the uncertain vector $w_k$ is bounded and $f_0(\cdot,\cdot)$ satisfies some Lipschitz assumptions, one can resort to bounded error methods \cite{MilaNortPieWal96} that guarantee that $y_k$ is always contained in $I(x_k)$. See, for example, \cite{MilaNovAUT05} and \cite{manzano2020robust}. Other bounded error strategies have been proposed in \cite{BaiTempo:99}, \cite{Jaulin00},  \cite{Bravo:2016:BoundingTechniques}. The statistical characterization of noise and disturbances can be used to enhance the performance of interval estimation me\-thods. See \cite{RollNazinLjung:05}, \cite{BravoAlamo:15}, \cite{Combastel15Aut} and references therein. Also, probabilistic validation methods can be used to assess the performance of the interval predictors \cite{Efron:86bootstrap}, \cite{alamo2015randomized}, \cite{alamo2018robust}, \cite{mirasierra2021prediction}.

Denote $F(\by |x_k)$ the cumulative distribution function of the associated output $y$ conditioned to the regressor $x=x_k$. That is,
$$ F(\by|x_k) = \Prob \set{ y \leq \by}{ x=x_k}. $$
Related with this probability is the notion of quantile  \cite{Murphy:12}, \cite{Koenker:1978:RegressionQuantiles}. Given $x_{k}$, we say that $\by_{\qp}$ is the conditioned $\qp$-quantile if
$$ F(\by_{\qp}|x_k) = \Prob \set{ y \leq \by_{\qp}}{ x=x_k} = \qp. $$
The notion of quantile is closely related to the one of confidence intervals.
The estimation of the conditioned quantiles is relevant in multiple applications (see \cite{Davino:14} and \cite{bassett2002portfolio}) and can be addressed using different methodologies.
The most classical approach relies on the assumption that $y_k$ and $x_k$ are jointly normal. That is, the assumption that the (joint) probability density function of the (random) variables $y$ and $x$ is a multivariable normal probability density function. Under this assumption, the conditioned p.d.f. is a monovariable normal p.d.f. and the quantiles can be obtained in a simple and direct way \cite{Papoulis:02}. Unfortunately, the methods based on normal distributions are very sensitive to the presence of outlier contamination. Moreover, in many long-tailed situations, the normal assumption is not well suited to characterize confidence intervals and one has to resort to non-Gaussian distributions. In these cases, generalizations of the Chebyshev inequality can be used to obtain probabilistic bounds \cite{navarro2016very}, \cite{stellato2017multivariate}. 

The computation of the conditioned quantiles can be also addressed by means of parametric regression techniques  \cite{Koenker:1978:RegressionQuantiles}, \cite{Davino:14}. If one assumes that there exists $\theta$ for which $y_k \approx \theta\T x_k$, then parameter vector $\theta$ can be chosen as the one that minimizes a cost function of the error $\theta\T x_k-y_k$. If one chooses a cost function that penalizes in an asymmetric way positive and negative errors then a quantile regressor is obtained. Given the training pairs $(y_j,x_j)$, $j=1,\ldots,N$ and $\qp\in (0,1)$, the quantile regressor is defined in terms of the following optimization problem
$$ \min\limits_{\theta} \Sum{j=1}{N} (1-\qp) \max\{0,\theta\T x_j-y_j\} + \qp\max\{0,y_j-\theta\T x_j\}.$$
This linear optimization problem penalizes the (training) errors $e_j=\theta\T x_j -y_j$, $j=1,\ldots,N$ in an asymmetric way. The positive errors are weighted with coefficient  $(1-\qp)$ and the negatives with coefficient $\qp$. If $\qp\in(0,1)$ is close to zero, then the positive errors will be highly penalized (in comparison with the negative ones). This means that every optimal solution $\theta_{\qp}$ to the linear optimization problem will tend to make most of the errors negative. This implies that  $\theta_{\qp}\T x_k$ could be used as a probabilistic lower bound for $y_k$. In a similar way, a probabilistic upper bound could be obtained taking $\qp\in(0,1)$ close to 1. Under rather mild assumptions, any minimizer $\theta_{\qp}$ of the proposed optimization problem can be used to obtain an estimation of the $\qp$ quantile. That is, $\theta_{\qp}\T x_k$ serves as an estimation of the $\qp$ quantile associated with $y_k$.
See \cite{Koenker:1978:RegressionQuantiles}, \cite{portnoy1997gaussian} and \cite{Davino:14}  for further details.

One of the main limitations of quantile regression is that a large number of training samples $N$ is required if one desires to obtain probabilistic guarantees of the method when $\qp$ is chosen close to the extremes of the interval $(0,1)$. This is due to the fact that estimating the probability of rare events requires a large number of samples. For example, the number of independent identically distributed samples required to obtain the $1-\epsilon$ quantile of a monovariable random variable grows with $\frac{1}{\epsilon}$ (see \cite{TeBaDa:97}, \cite{alamo2015randomized} and \cite{alamo2018robust}).

This paper presents a new methodology for the computation of interval predictions of a dynamical system. Dissimilarity functions are used to estimate the conditional probability density function of the outputs. The estimated probability density function is used to derive the interval prediction. It is shown that the standard linear regression is a particular case of the proposed methodology. The paper is organized as follows. In Section \ref{sec:DissimilarityFunctions} a family of dissimilarity functions is proposed. In Section \ref{sec:regression} the role of dissimilarity functions in linear regression is analyzed. The probabilistic interval predictors are presented in Section \ref{sec:IntervalEstimation}. The methodology is applied to some forecasting problems in Section \ref{sec:example}. The paper ends with a section of conclusions.

\section{Dissimilarity functions}\label{sec:DissimilarityFunctions}

Given a data set
$$ \cD=\set{z_i}{i=1,\ldots,N} \subset \R^n,$$
 we are interested in determining if a given vector $z$ can be considered to be similar to the other vectors of the data set $\cD$.
 In a more precise way, we are looking for a function
 $$J_d(\cdot,\cdot):\R^n\times \cD\to [0,\infty]$$
  that measures the dissimilarity between a given point $z$ and the data set $\cD$. Large values of $J_d(z,\cD)$ represent a high degree of dissimilarity, while small values correspond to a high degree of similarity (i.e., a small degree of dissimilarity).
 Clearly, from a dissimilarity function $J_d(x,\cD)$ one can obtain a similarity function $J_s(x,\cD)$. For example, given $\sigma>0$, $J_s(x,\cD)=\mathrm{e}^{-\sigma J_d(z,\cD)}$ is small when $z$ is not similar to the points in $\cD$ and close to $1$ when $z$ is very similar to the elements of $\cD$. Another possibility would be $J_s(x,\cD)=(1+\sigma J_d(z,\cD))^{-1}$, where $\sigma>0$.

 There exists a wide class of operators that can serve as dissimilarity functions for the particular case in which $\cD$ is a singleton ($\cD=\{z_\cD\}$). For singleton $\cD$, one popular choice is
 $$ J_d(z,z_\cD) = \|z-z_\cD\|,$$
 where $\|\cdot\|$ is a given norm. One could also use the minimum distance to set $\cD$. That is,
   \begin{equation}\label{equ:min:distance}
    J_d(z,\cD) = \min\limits_{\hat{z}\in \cD} \, \|z-\hat{z}\|.\end{equation}
   Another possibility could be to consider as a dissimilarity function the mean value of the distances of $z$ to each member of set $\cD$. See chapter 2 of \cite{Goshtasby:12} and chapter 2 of \cite{wierzchon2018modern} for a review of similarity and dissimilarity functions applied in the field of image registration and in the context of cluster analysis, respectively.

 Dissimilarity and similarity functions can be used in the context of regression.  Suppose that we have the pairs $\{x_i,y_i\}$, $i=1,\ldots,N$ and that we would like to estimate, given $x$, its corresponding output $y$. Given the similarity function $J_s(\cdot,\cdot)$, one possibility for the estimation $\hat{y}$ of $y$ is  $$ \hat{y} = \Sum{i=1}{N} \lambda_i y_i, $$
 where the scalars $\lambda_i$ are chosen in such a way that $\lambda_i$ is small when the similarity function $J_s(x,x_i)$ is small.
 It is also reasonable to normalize the sum of the scalars $\lambda_i$  to the unity. That is, $\Sum{i=1}{N} \lambda_i=1$.  For example, one could choose $$ \lambda_i = \frac{J_s(x,x_i)}{\Sum{j=1}{N} J_s(x,x_j)}, \;\; i=1,\ldots,N.$$
 Although this approach could be valid for some applications, more sophisticated approaches are required in many situations, as it is just a weighted average.  We propose in this paper a convex optimization problem to obtain a measure of dissimilarity between a point $z$ and a set $\cD$. This is formally stated in the following definition.

\begin{definition}\label{def1}
Given $z\in \R^n$, a set of measurements $\cD=\{z_1,\ldots,z_N\}\subset \R^n$ and the scalar $\gamma\geq 0$, the dissimilarity function $J_{\gamma}(z,\cD)$ is defined as
\begin{eqnarray}
J_{\gamma}(z,\cD)&=&\min\limits_{\lambda_1,\ldots,\lambda_N}  \Sum{i=1}{N} \lambda_i^2 + \gamma \Sum{i=1}{N}|\lambda_i|  \nonumber \\
s.t. && z=\Sum{i=1}{N}\lambda_i z_i \nonumber\\
&& 1 = \Sum{i=1}{N}\lambda_i \label{problem:general}.
\end{eqnarray}
\end{definition}

\begin{remark} Note that non negative constant weights could be included into the cost function. That is, one could consider the cost function
$$ \Sum{i=1}{N} w_i\lambda_i^2 + \gamma \Sum{i=1}{N}|\lambda_i|,$$
where the scalars $w_i$, $i=1,\ldots,N$ are used to weight the different elements in $\cD$. These weights could be computed using a distance function between $z$ and the singleton $\{z_i\}$ (for example,  $w_i = ||z-z_i||$) or any dissimilarity function. 

This would be a way to incorporate local information into the analysis. This strategy could be useful when the considered system is non-linear. Although the results of the paper are stated for the particular case in which $w_i=1$, $i=1,\ldots,N$, the generalization to the general case is not difficult.
\end{remark}

\begin{remark}
We notice that the optimization problem (\ref{problem:general}) could be non-feasible. In order to rule out this possibility, we assume that the vectors that compose set $\cD$ span all the space.
\end{remark}

\begin{remark}
Optimization problem (\ref{problem:general}) is similar to the one appearing in the context of direct weight optimization and kriging, where central predictions of a certain variable are obtained by means of the solution of an optimization problem \cite{RollNazinLjung:05}, \cite{Bravo:2016:BoundingTechniques}, \cite{salvador2019offset}, \cite{cressie1986kriging}, \cite{salvador2018data}.
\end{remark}

It is important to remark that the proposed dissimilarity measure is invariant with respect to affine transformations. This is formally stated in the following property.

\begin{property}
Consider $z_{T,v}$ and $\mathcal{D}_{T,v}$ obtained from $z$ and $\mathcal{D}$ through the following affine transformation.
\begin{eqnarray*} z_{T,v} &=& Tz + v \\
\mathcal{D}_{T,v} &=& \set{Tz+v}{z\in \mathcal{D}},
\end{eqnarray*}
where $T$ is any non-singular matrix and $v$ is any vector of adequate dimensions. Then
$$ J_{\gamma}(z,\mathcal{D}) = J_{\gamma}(z_{T,v},\mathcal{D}_{T,v}).$$
\end{property}

\proof
We first show that any feasible solution $\lambda_i$, $i=1,\ldots,N$ to the problem of computing $J_{\gamma}(z,\mathcal{D})$ is also a feasible solution for the computation of $J_{\gamma}(z_{T,v},\mathcal{D}_{T,v})$. Suppose that $z=\Sum{i=1}{N} \lambda_i z_i$ and $\Sum{i=1}{N} \lambda_i=1$. Then
\begin{eqnarray*}
z_{T,v} & = & Tz+v \\
& = & T\left( \Sum{i=1}{N} \lambda_i z_i\right) +  \left( \Sum{i=1}{N} \lambda_i \right)v \\
& = & \Sum{i=1}{N} \lambda_i (Tz_i + v).
\end{eqnarray*}
We notice that $Tz_i+v$, $i=1,\ldots,N$, are the elements of $\mathcal{D}_{T,v}$. Therefore $\lambda_i$, $i=1,\ldots,N$, is also a feasible solution for the problem that defines $J_{\gamma}(z_{T,v},\mathcal{D}_{T,v})$. From this we infer that $J_{\gamma}(z_{T,v},\mathcal{D}_{T,v})\leq J_{\gamma}(z,\mathcal{D})$. On the other hand, since $T$ is non-singular we can make a similar reasoning to show that any feasible solution for $J_{\gamma}(z_{T,v},\mathcal{D}_{T,v})$ is a feasible solution for $J_{\gamma}(z,\mathcal{D})$. In this way we prove also that $J_{\gamma}(z,\mathcal{D})\leq J_{\gamma}(z_{T,v},\mathcal{D}_{T,v})$. Both inequalities prove the claimed  equality.
\QED

This invariance property is very important because it gua\-ran\-tees that the analysis based on the proposed dissimilarity function is not affected by the choice of the coordinate system. We notice that many of the dissimilarity functions that can be found in the literature are not invariant. For example, any dissimilarity function based on the distance of $z$ to the elements of $\mathcal{D}$, such as that of equation (\ref{equ:min:distance}),  will be dependent on the particular choice of coordinate system. 

The proposed optimization problem (\ref{problem:general}) is a strict convex optimization problem subject to convex constraints. This means that
 it has a unique solution \cite{Boyd04}. From an optimization point of view, we notice that the numerical resolution can be addressed using a dual formulation.  In the dual formulation for this particular optimization problem, the number of dual decision variables is equal to the number of equality constraints ($n+1$) which is in many situations much smaller than the number of primal variables ($N$). On the other hand, the gradient of the objective function in the dual formulation can be obtained in a direct way because once the dual variables are fixed, the optimal values for the primal variables are obtained solving a separable optimization problem (which has an explicit solution). The numerical examples of this paper have been obtained using an accelerated gradient method in the dual variables. See \cite{beck2017first}, \cite{Beck09}  and \cite{nesterov2018lectures}. The alternating direction method of multipliers can also be used in this context \cite{Boyd10}.

 As it is formally stated in the following property, the optimization problem has an explicit solution for the particular case $\gamma=0$ (see Appendix A for proof). 

\begin{property}\label{property:explicit:gamma:zero} Suppose that $\mathcal{D}=\{z_1,z_2,\ldots,z_N\}$, then $J_{0}(z,\mathcal{D})$ has the following explicit expression
 $$J_{0}(z,\mathcal{D}) = N^{-1} + (z-\bar{z})\T(ZZ\T-N\bar{z}\bar{z}\T)^{-1}(z-\bar{z}),$$ where $Z = [ z_1\;z_2\;...\;z_N ]$, $\bar{z} = N^{-1}Zu$ and $u \in \R^N$ is a vector with all its $N$ components equal to 1. 
\end{property}

The previous result shows that the dissimilarity function is a quadratic function on the argument $z$ for the particular case $\gamma=0$.
For the more general case in which $\gamma>0$ we can infer from the Karush-Kuhn-Tucker optimality conditions \cite{Boyd04} that the dissimilarity function $J_{\gamma}(z,\mathcal{D})$ is a piecewise convex quadratic function with respect to $z$.

\section{Dissimilarity functions and regression}\label{sec:regression}

We show in this section how dissimilarity functions can be used in the context of regression. Imagine that the data set $$ \mathcal{D} = \set{z_i = \bv y_i \\ x_i \ev}{ i=1,\ldots,N}\subset \cY\times \cX,$$ is available.  Given $x_k$, and $\gamma\geq 0$, one could obtain and estimation $\hat{y}_k$ for $y_k$ minimizing the dissimilarity function of vector $\bv y \\ x_k\ev$ with respect to the data set $\cD$.
  That is,
  $$ \hat{y}_k = \arg\,\min\limits_y J_{\gamma}(\bv y\\ x_k\ev,\mathcal{D}). $$

Therefore, given $x_k$ and $\gamma\geq 0$, the estimation $\hat{y}_k$ could be obtained from the optimization problem

\begin{eqnarray}
\min\limits_{y,\lambda_1,\ldots,\lambda_N}  && \Sum{i=1}{N} \lambda_i^2 + \gamma \Sum{i=1}{N}|\lambda_i|  \nonumber \\
s.t. && \bv y \\ x_k \ev =\Sum{i=1}{N}\lambda_i \bv y_i \\ x_i\ev \label{equ:y:lambda} \\
&& 1 = \Sum{i=1}{N}\lambda_i. \nonumber
\end{eqnarray}

Since the decision variable $y$ appears only in the equality constraint (\ref{equ:y:lambda}), one could solve the optimization problem ignoring the equality constraint
$$ y = \Sum{i=1}{N} \lambda_i y_i, $$
and making the optimal value of $y$ equal to
\begin{equation} \label{equ:optimal:y}
\hat{y}_k = y^* = \Sum{i=1}{N} \lambda_i^* y_i,
\end{equation}
where $\lambda_i^*$, $i=1,\ldots,N$ are the optimal values of the optimization problem

\begin{eqnarray*}
\min\limits_{\lambda_1,\ldots,\lambda_N}  && \Sum{i=1}{N} \lambda_i^2 + \gamma \Sum{i=1}{N}|\lambda_i|  \\
s.t. && x_k =\Sum{i=1}{N}\lambda_i x_i  \\
&& 1 = \Sum{i=1}{N}\lambda_i.
\end{eqnarray*}

 Therefore, the estimation provided by the method is a linear combination of the outputs $y_i$. This is consistent with different results from the specialized literature in which it is shown that under certain assumptions, the optimal solution to an estimation problem is given by a linear combination of the observed outputs (see \cite{RollNazinLjung:05}, and \cite{MilaBel82}). The central estimation provided by equation (\ref{equ:optimal:y}) is similar to other weighted methods like \cite{Bravo:2016:BoundingTechniques}, \cite{RollNazinLjung:05} and \cite{salvador2019offset}.

As it is stated in the following property, the estimation obtained for the particular case $\gamma=0$ matches the one given by the linear least squares method.
\begin{property} Given a point $x_k$, the estimation
$$\hat{y}_k = \arg\,\min_{y}\; J_{0}(\bv y \\ x_k \ev,\mathcal{D}),$$  matches the estimation obtained by linear least squares using $\mathcal{D} = \set{z_i =\bv y_i \\ x_i \ev}{ i=1,\ldots,N}$ as data set. The proof is provided in Appendix B.
\end{property}

Previous property shows that the estimation method proposed in this paper encompasses the least squares method for the particular case $\gamma=0$.
A family of optimal estimators is obtained if one considers $\gamma$ as a tuning parameter. In the following sections, we show not only how to tune the value of $\gamma$, but also how to use this methodology to obtain probabilistic interval estimations.

\subsection{Empirical Probability Density Function}

The dissimilarity of a given vector $z\in \cal{Z}$ $\subseteq \R^{n}$ with respect to the elements of $\mathcal{D}$ can be used to define an empirical probability density function. The next definition introduces the notion of empirical p.d.f.

\begin{definition}[Empirical p.d.f.] Given a set of measurements $\mathcal{D}=\{z_1,\ldots,z_N\}\subset\mathcal{Z}$, $\gamma\geq 0$ and $c\geq 0$, the empirical probability density function
$\ep_{\gamma,c}(z,\mathcal{D})$ is defined for every $z\in \cal{Z}$ as
\begin{equation}\label{equ:epdf}\ep_{\gamma,c}(z,\mathcal{D})= \frac{\exp\left(-c J_{\gamma}(z,\mathcal{D})\right)} {\int\limits_{\mathcal{Z}}\exp\left(-c J_{\gamma}(\hat{z},\mathcal{D})\right)d\hat{z}}. \end{equation}
\end{definition}
Note that expression (\ref{equ:epdf}) provides a family of probability density functions that are parameterized by constants $\gamma$ and $c$.
We notice that by construction,
$$ \int\limits_{\mathcal{Z}}\ep_{\gamma,c}(\hat{z},\mathcal{D})d\hat{z} = 1.$$
We also notice that if $\mathcal{Z}$ is a compact set, the choice $c=0$ provides a uniform p.d.f. in $\mathcal{Z}$.
\begin{remark} Recall that from property \ref{property:explicit:gamma:zero} we have
\begin{eqnarray*}
J_{0}(z,\mathcal{D}) &=&  N^{-1} + (z-\bar{z})\T(ZZ\T-N\bar{z}\bar{z}\T)^{-1}(z-\bar{z}) \\
&=&  \frac{1}{N} + \frac{1}{N}(z-\bar{z})\T(\frac{1}{N} ZZ\T-\bar{z}\bar{z}\T)^{-1}(z-\bar{z}),
\end{eqnarray*}
where $Z = [ z_1\;z_2\;...\;z_N ]$, $\bar{z} = N^{-1}Zu$ and $u$ is a vector with all its components equal to 1. This means that if $c = \frac{N}{2}$ and $\gamma=0$ then $\ep_{\gamma,c}(z,\mathcal{D})$ is a multivariable normal distribution with mean $\bar{z}$ and covariance matrix $ \frac{1}{N} ZZ\T-\bar{z}\bar{z}\T$, which corresponds to the empirical covariance matrix of the data set $\mathcal{D}$.
\end{remark}
As already commented, the proposed method provides a way to obtain a family of empirical probability density functions that encompasses the normal distributions and the uniform one. In order to obtain the parameters $c$ and $\gamma$ for a given data set $\mathcal{D}$ generated by other distributions, one could use the maximum likelihood methodology. See, for example, \cite{Murphy:12}. We show in the following example how the proposed methodology  can be used to estimate the probability density function that characterizes a given data set $\mathcal{D}$.

\subsection{Clarifying example}
\label{sec:clarifying}
A sample of 600 points in $\R$ is obtained from a uniform probability function with support $[0,1]$. One half of the available points are used as set $\mathcal{D}$. The other half is used as a test set. Then for every point in the test set, equation (\ref{equ:epdf}) is used  to estimate the empirical probability density function associated with the considered pairs $(c,\gamma)$. 

\begin{figure}
\centering
\includegraphics [width=85mm,height=60mm]{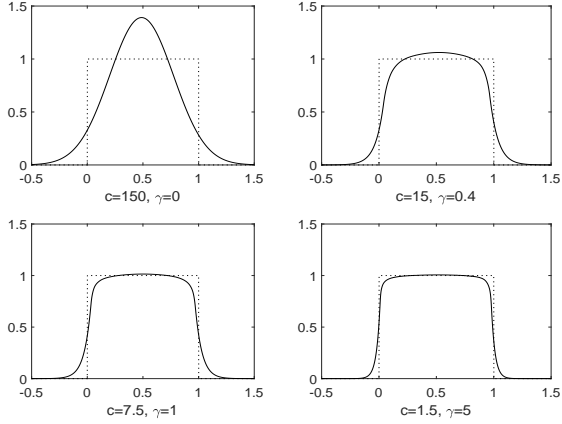}
\caption{Estimated probability distribution functions.}\label{DistRef2}
\end{figure}

Figure \ref{DistRef2} shows the empirical probability density functions estimated using different values of the parameters $c$ and $\gamma$. In this case, $c=1.5$ and $\gamma=5$ is the pair that achieves the best fit for the distribution proposed in this example. 

\section{Interval estimation}\label{sec:IntervalEstimation}

This section presents the methodology to obtain, given $x_k \in \mathcal{X}$,  an interval estimation of the corresponding output $y_k$.  
Given $x_k\in \mathcal{X}$, the data set 
$$ \mathcal{D} = \set{\bv y_i \\ x_i \ev}{ i=1,\ldots,N} \subset \mathcal{Y}\times\mathcal{X},$$ and the non negative scalars $c$ and $\gamma$, the empirical {\bf conditional} p.d.f. in $\mathcal{Y} \times \mathcal{X}$ is defined as 
$$
    \ecp_{\gamma, c}(y,x_k,\mathcal{D}) = \frac{\exp\left(-c J_{\gamma}(\bv y \\ x_k \ev, \mathcal{D} )\right)}{\int\limits_{\mathcal{Y}} \exp\left(-c J_\gamma(\bv \hat{y} \\ x_k \ev, \mathcal{D})\right)d\hat{y}}, \; \forall y\in \mathcal{Y}. $$
The empirical conditional p.d.f. serves to model the probability of $y$ given the occurrence of $x_k$. We now show how to use this notion to compute, given $x_k$, an interval estimation of $y_k$.

First, in order to simplify the numerical integration required to compute the interval estimations, we approximate set $\cY$ with a set  $\bar{\cY}$ of finite cardinality. That is, we consider the set 
$$ \bar{\cY}=\{\by_1, \ldots, \by_M \},$$
where $\by_j< \by_{j+1}$, $j=1,\ldots,M-1$, and the extreme values of $\bar{\cY}$ are chosen to guarantee that $y_k$ belongs to $[\by_1,\by_M]$ with high probability. A reasonable procedure to construct set $\bar{\cY}$ is to define $\by_1,\ldots,\by_M$ as follows
\begin{eqnarray*}
\by_1&=&\min\limits_{i=1,\ldots,N} y_i\\
\by_M&=&\max\limits_{i=1,\ldots,N} y_i\\ 
\by_j&=& \by_1+ \left(\frac{\by_M-\by_1}{M-1}\right)(j-1),  \; j=2,\ldots,M-1.
\end{eqnarray*}
We notice that larger values of $M$ provide better approximations of $\cY$ at the expense of a larger computational burden.
Given $x_k\in \cX$, we define the \textit{discrete} empirical conditional distribution, defined at each point of $\bar{\cY}$, as 

\begin{equation}
    \becp_{\gamma, c}(y,x_k,\mathcal{D}) = \frac{\exp\left(-c J_{\gamma}(\bv y \\ x_k \ev, \mathcal{D} )\right)}{\Sum{j=1}{M} \exp\left(-c J_\gamma(\bv \by_j \\ x_k \ev, \mathcal{D})\right)}. \label{eq:decp}
\end{equation}
 By construction, 
 $$\Sum{\ell=1}{M} \becp_{\gamma, c}(\by_\ell,x_k,\mathcal{D})=1.$$
This discrete empirical conditional p.d.f. defines a conditioned probability distribution on $\bar{\cY}$ (given $x_k$), that we denote  $\Prob_{\bar{\cY}|x_k}$. According to this discrete distribution, we have that $\by_\ell$ (i.e. the $\ell$-th element of $\bar{\cY}$) satisfies,\begin{eqnarray} \Prob_{\bar{\cY}|x_k} \, \{y \leq \bar{y}_\ell\} &=& \Sum{j=1}{\ell} \becp_{\gamma, c}(\by_j,x_k,\mathcal{D}), \label{equ:prob:lower} \\ \Prob_{\bar{\cY}|x_k} \, \{y \geq \bar{y}_\ell\} &=& \Sum{j=\ell}{M}\becp_{\gamma, c}(\by_j,x_k,\mathcal{D}). \label{equ:prob:upper}
\end{eqnarray}
Given $x_k$, $\gamma\geq 0$ and $c\geq 0$, and $\qp\in(0,1)$, we define the empirical upper conditioned $\qp$-quantile, denoted by  $y_{\qp}^+$, as the \textit{smallest} element of $\bar{\cY}$ that satisfies
$$ \Prob_{\bar{\cY}|x_k} \,\{ y \leq y_{\qp}^+\} \geq 1-\qp. $$
In a similar way, the empirical lower conditional $\qp$-quantile,  denoted by $y_\qp^-$,
is defined as the \textit{largest} element of $\bar{\cY}$ that satisfies $$ \Prob_{\bar{\cY}|x_k} \,\{ y \geq y_{\qp}^-\} \geq 1-\qp. $$

Given $x_k$ and $\qp\in(0,1)$, the interval prediction for $y_k$ is $[y_\qp^-,y_\qp^+]$. 
According to the discrete distribution $\Prob_{\bar{\cY}|x_k}$, and the definition of $y_\qp^-$ and $y_\qp^+$ we have 
$$ \Prob_{\bar{\cY}|x_k}\{y\in [y^-_\qp,y^+_\qp] \} \geq  1 - 2\qp.$$
What precedes illustrates how to compute the interval prediction $[y_\qp^-,y_\qp^+]$, for a given $x_k$, $\qp$ and pair ($\gamma,c$). See  Algorithm \ref{alg:ComputationInterval} for a detailed description of the procedure.

\begin{remark}[Conditioned median]\label{remark:median}  Given the occurrence of $x_k$, a sensible estimation for $y_k$ is the conditioned median $y_m(x_k,\gamma,c)$, that can be approximated by the center of the interval $[y^-_{0.5},y^+_{0.5}]$.
\end{remark}

\begin{algorithm}[h]
 \caption{Interval estimation $[y_\qp^-(x,\gamma,c), y_\qp^+(x,\gamma,c) ]$.  \label{alg:ComputationInterval}}
\begin{algorithmic}[1] 

\REQUIRE $x$, $\qp\in(0,1)$, $\gamma\geq 0$, $c\geq 0$, $\cD$, $\bar{\cY}=\{\by_1,\ldots, \by_M\}$.
\ENSURE $y_\qp^-$, $ y_\qp^+$. 

\STATE  Obtain the dissimilarity function (see Definition \ref{def1}) for each element of $\bar{\cY}$:  
$$ d_j = J_{\gamma}\left(\bmat{c} \by_j \\ x    \emat , \mathcal{D} \right), \; j=1,\ldots,M.$$
\STATE Compute the conditioned probabilities (see equation \ref{eq:decp}): 
$$ p_j = \becp_{\gamma,c}(\by_j,x,\cD)= \frac{\exp\left( -cd_j\right)}{\Sum{\ell=1}{M} \exp\left( -cd_\ell\right) }, \; j=1,\ldots,M.$$ 

\STATE Compute the indexes $\ell_\qp^+$ and $\ell_\qp^-$ corresponding to the lower and upper conditioned $\qp$-quantiles (see (\ref{equ:prob:lower}) and (\ref{equ:prob:upper})):
\begin{eqnarray*}
\ell_\qp^+&=&\mbox{smallest integer $\ell$ satisfying } \Sum{ j=1 }{\ell} p_j \geq 1-\qp,\\
\ell_\qp^-&=&\mbox{largest integer $\ell$ satisfying } \Sum{j=\ell}{M} p_j \geq 1-\qp.\\
\end{eqnarray*}
\STATE $y^-_\qp=\by_{\ell_\qp^-}$ and $y^+_\qp=\by_{\ell_\qp^+}$.

\end{algorithmic}
\end{algorithm}

The properties of the prediction intervals obtained using the procedure detailed above rely on the specific choice for $\gamma$ and $c$, since they determine the underlining empirical distribution (see the example of section \ref{sec:clarifying}). Given $\qp\in(0,1)$, we now detail how to obtain a pair $(\gamma^*_\qp,c^*_\qp)$ such that sharp interval estimations are obtained, while meeting the probabilistic specifications (determined by $\qp$) in a validation set 
$$ \cV = \set{\bv \ty_s \\ \tx_s \ev}{ s=1,\ldots,N_{\cV}} \subset \cY\times\cX.$$

Let us now analyze the role of parameter $c\geq 0$ in the discrete empirical conditioned distribution given in equation (\ref{eq:decp}). On the one hand, the choice $c=0$ provides a flat distribution in which each element of $\bar{\cY}$ has a conditioned probability equal to $\frac{1}{M}$. On the other hand, large values of $c$ provide narrow distributions centered around the point in $\bar{\cY}$ that minimizes, given $x_k$, the dissimilarity function $J_\gamma(\cdot,\cD)$. Consequently, for a fixed value of $\gamma$, larger values of $c$ reduce the size of the obtained interval at the  expense of increasing the fraction of outputs that are not contained in the interval estimations. Therefore, given $\gamma$, the corresponding value for $c$ should be chosen as the largest value of $c$ that guarantees in the validation set that the obtained intervals contain the outputs with the desired probability.

\begin{algorithm}[h]
 \caption{Optimal value of $c\geq 0$, for given $\gamma\geq 0$ and $\tau\in(0,1)$ \label{alg:computation:c}}

\begin{algorithmic}[1] 

\REQUIRE $\qp\geq 0$, $\gamma\geq 0$, $c_{\max}>0$ and $\epsilon>0$, $\cD$, $\bar{\cY}$ and 
the validation data set $\cV =  \set{\bv \ty_s \\ \tx_s \ev}{ s=1,\ldots,N_{\cV}} \subset \cY\times\cX$.
\ENSURE $c_\gamma$.
\STATE $c_{\min}=0$.
\WHILE { $c_{\max}-c_{\min} \geq \epsilon$} 
\STATE $c=\frac{1}{2}(c_{\max}+c_{\min})$.
\STATE Compute, using Algorithm \ref{alg:ComputationInterval}, the $N_\cV$ intervals 
$$ I_s=[y_\qp^-(\tx_s,\gamma,c), y_\qp^+(\tx_s,\gamma,c)], \; s=1,\ldots,N_\cV. $$  
\STATE Make $n^+_{\rm{viol}}$ equal to the number of violations of the upper constraints 
$$ \ty_s \leq y_\qp^+(\tx_s,\gamma,c), \; s=1,\ldots,N_\cV,$$
and $n^-_{\rm{viol}}$ equal to the number of violations of the lower constraints
$$\ty_s \geq y_\qp^-(\tx_s,\gamma,c), \; s=1,\ldots,N_\cV.$$
\IF{ $ \fracg{\max\{n^+_{\rm{viol}},n^-_{\rm{viol}}\}}{N_\cV}< \tau $}

    \State $c_{\min}=c$,
\ELSE{}
    \State $c_{\max}=c$.
\ENDIF
\ENDWHILE
\STATE $c_\gamma=c_{\min}$.
\end{algorithmic}
\end{algorithm}

From the discussion above, we have that the parameter $c$ corresponding to a particular choice of $\gamma>0$ (denoted $c_\gamma$) is determined by $\tau$. As it is detailed in Algorithm \ref{alg:computation:c}, $c_\gamma$ is chosen as the largest value of $c$ (up to a given accuracy $\epsilon>0$) that guarantees in the validation set that the obtained confidence intervals contain the outputs with the desired probability. That is, non smaller than $1-2\tau$.

Parameter $\gamma>0$ can be obtained maximizing the likelihood ratio which, for a specific $\gamma$ and corresponding $c_\gamma$, is defined as
$$ 
L_{\gamma} = \sum_{s=1}^{N_\cV} \textrm{log}\left( \textrm{ecp}_{\gamma,c_\gamma} (\ty_{j},\tx_{j},\mathcal{D}) \right)
$$
Using $\bar{\cY}=\{\by_1,\ldots,\by_M\}$, a numerical approximation to the optimal value of $\gamma$ is given by 
\begin{equation}
\gamma_\qp^*    \approx  \arg\max_{\gamma\in \Gamma }\sum_{s=1}^{N_\cV} \textrm{log}\left( \frac{\exp\left(-c_\gamma J_{\gamma}(\bv \ty_s \\ \tx_s \ev, \mathcal{D} )\right)}{\Sum{j=1}{M} \exp\left(-c_\gamma J_\gamma(\bv \by_j \\ \tx_s \ev, \mathcal{D})\right)}  \right),
  \label{eq:max_lik}
  \end{equation}
  where $\Gamma$ is a set containing all the possible values considered for $\gamma$.

\begin{remark}
Other criteria can be used to compute $\gamma^*_\qp$. For example, $\gamma^*_\qp$ could be obtained by minimizing a cost function $Q_{\gamma}$ that penalizes the average length of the intervals and/or the average prediction error with respect to the conditioned median introduced in Remark \ref{remark:median}. 
We notice, however, that explicitly minimizing the size of the intervals may translate into an increased violation rate when the validation set has not a sufficiently large number of samples.  
\end{remark}

\section{Example} \label{sec:example}

\begin{table*}[htbp]
\caption{Results for the Lorenz Attractor, interval $[o_{5\%},o_{95\%}]$.}
\label{tab:tabLorenz90}
\centering

\begin{tabular}{c|cc|cc|cc}
\hline
 & \multicolumn{2}{|c|}{\textbf{Proposed approach}}  & \multicolumn{2}{|c}{\textbf{Quantile Regression}} & \multicolumn{2}{|c}{\textbf{Set Membership}} \\
\hline
Data set length & \textbf{Empirical Probability}  & \textbf{Interval Width} & \textbf{Empirical Probability} & \textbf{Interval Width} & \textbf{Empirical Probability} & \textbf{Interval Width}   \\
\hline
200 & 0.9140  & 2.0578  & 0.8290 & 3.0965 & 0.8960 & 2.9378 \\
350 & 0.8990  & 1.9352  & 0.8260 & 3.0550 & 0.9100 & 2.4773 \\
500 & 0.9070  & 2.0223  & 0.8410 & 3.2450 & 0.9120 & 2.5671\\
\hline 
\end{tabular}
\end{table*}

\begin{table*}[htbp]
\caption{Results for the Lorenz Attractor, interval $[o_{10\%},o_{90\%}]$.}
\label{tab:tabLorenz80}
\centering

\begin{tabular}{c|cc|cc|cc}
\hline
 & \multicolumn{2}{|c|}{\textbf{Proposed approach}}  & \multicolumn{2}{|c}{\textbf{Quantile Regression}} & \multicolumn{2}{|c}{\textbf{Set Membership}} \\
\hline
Data set length & \textbf{Empirical Probability}  & \textbf{Interval Width} & \textbf{Empirical Probability} & \textbf{Interval Width} & \textbf{Empirical Probability} & \textbf{Interval Width}   \\
\hline
200 & 0.8060  & 1.6053  & 0.7450 & 2.2776 & 0.8160 & 2.4248 \\
350 & 0.8060  & 1.6164  & 0.7270 & 2.0607 & 0.7900 & 1.9797 \\
500 & 0.8100  & 1.6195  & 0.7630 & 2.4371 & 0.8100 & 2.0021\\
\hline 
\end{tabular}
\end{table*}

The Lorenz attractor is a system of ODEs known for having chaotic solutions with certain values of the parameters of the system. 
The equations that define the system are the following
\begin{align}
    &\frac{\textrm{d}o}{\textrm{d}t} \, = \, \sigma (p-o) \nonumber \\
    &\frac{\textrm{d}p}{\textrm{d}t} \,=\, o (\rho-q) - p \\
    &\frac{\textrm{d}q}{\textrm{d}t} \,=\, op - \beta q\,, \nonumber
\end{align}
where $\sigma$, $\rho$ and $\beta$ are real scalar parameters. In this example, these parameters take the values $\sigma=10$, $\rho=28$ and $\beta=8/3$. Furthermore, in order to obtain the necessary data, the  ODEs have been integrated numerically with a fixed time step of $T_s = 0.1s$ and  initial conditions $o(0)=1$, $p(0)=1$ and $q(0)=1$.   Here, it is considered the task of forecasting the one-step ahead value of $o$, i.e., $y_{k} = o_{k}$, using the two previous values of $o$, that is,  the regressor vector will be $x_k = [y_{k-1}, \,y_{k-2}]\T$. To start with, $2500$ data points are considered, normalized in the $[0,1]$ range. Different sizes for the data set $\mathcal{D}$ are considered in this example ($200$, $350$ and $500$ points). The validation set $\cV$  consists of $1000$ data points and other $1000$ data points are used as a test set, denoted by $\mathcal{S}$ (note that $\mathcal{D}$, $\cV$ and $\mathcal{S}$ are mutually disjoint sets). The set $\Gamma$ is taken from $[0,3]$ using a $0.1$ sampling step. On the other hand, $\mathcal{\bar{Y}}$ is obtained from a grid of equally distant points in the interval $[-0.1893,1.2298]$ sampled with a $1.4191\times 10^{-4}$ step. 

Two different techniques will be considered as benchmarks. The first one is quantile regression \cite{Koenker:1978:RegressionQuantiles}, \cite{Davino:14}, a classical method for the estimation of  conditioned quantiles. The second one is the set-membership method described in \cite{milanese2004set,milanese2011unified}. This technique is a well-known method to generate interval bounds for a time series (usually produced from a dynamical system). For the sake of comparison, to guarantee that these bounds contain the output within a prescribed probability, we choose the parameters $\epsilon, \gamma$ of \cite{milanese2004set} such that the resulting empirical probability of containing a sample  within the validation set $\cV$ is no smaller than $1-2\tau$.

\begin{figure}[h]
    \centering
    \includegraphics[width=0.45\textwidth]{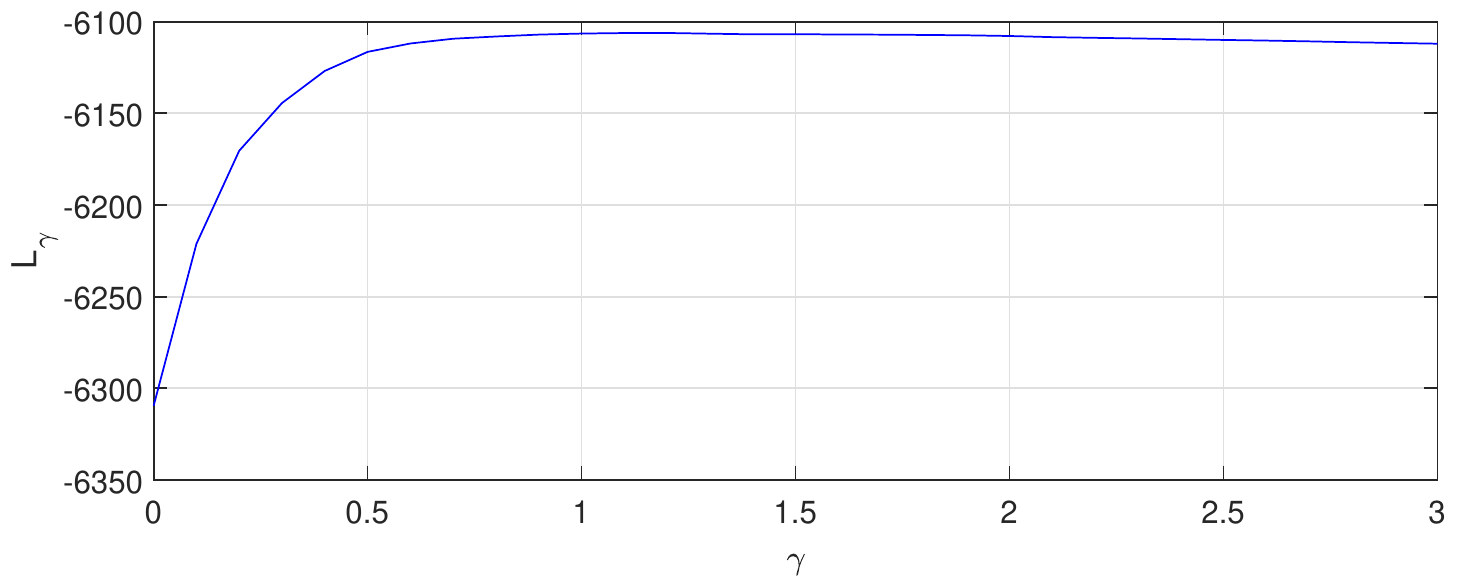}
    \caption{Maximum likelihood ratio as a function of $\gamma$.}
    \label{fig:max_lik}
\end{figure}

 The numerical results of the proposed approach and the two benchmark techniques are shown in table \ref{tab:tabLorenz90} for a $[o_{5\%},o_{95\%}]$ interval, that is, $\qp=0.05$, and in table \ref{tab:tabLorenz80} for $[o_{10\%},o_{90\%}]$ ($\qp=0.1$). The output of the test data should be contained in the first interval with a probability of $0.9$ ($0.8$ for the second interval). The optimal value for $\gamma$ has been chosen maximizing the likelihood function $L_\gamma$ (see equation (\ref{eq:max_lik}) and figure \ref{fig:max_lik}). 
 
 The empirical probability in the case of the quantile regression clearly does not meet the probabilistic specifications. In the case of the proposed approach and the set-membership method, the observed fraction of outputs that fall into the predicted intervals is much closer to the desired one. Note that, for all techniques, the obtained empirical probability can be below the desired probability. This could be solved relying on a probabilistic scaling scheme \cite{mirasierra2021prediction} or on probabilistic validation schemes \cite{carnerero2021probabilistically}, \cite{karg2019probabilistic}.

 Regarding the interval width, the proposed approach clearly manages to obtain the smallest intervals for each data set. 
 For the $[o_{5\%},o_{95\%}]$ interval,   the interval width is on average $24.35\%$ smaller than those provided by the set-membership method and $35.96\%$  than the quantile regression. On the other hand, for the $[o_{10\%},o_{90\%}]$ interval, the intervals with the proposed technique are $23.75\%$ smaller than those with set-membership and $28.21\%$ than those with the quantile regression.  Taking into account the empirical probability values and the interval widths, we can conclude that the proposed approach obtains the best results.
 
\begin{figure}[h]
    \centering
    \includegraphics[width=0.45\textwidth]{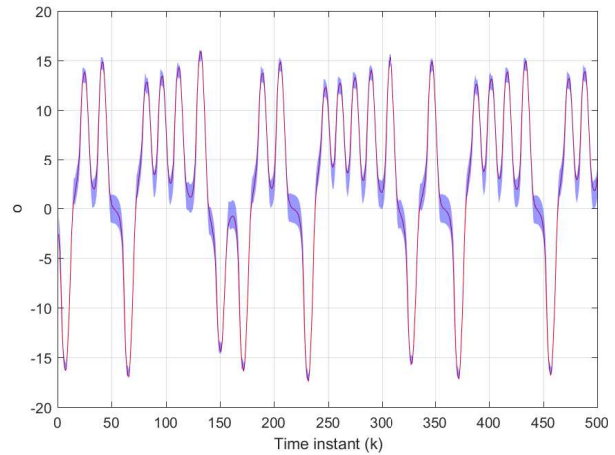}
    \caption{Test set and computed intervals for Lorenz Attractor (interval $[o_{5\%},o_{95\%}]$).}
    \label{fig:lorenz_90}
\end{figure}

Finally, in figure \ref{fig:lorenz_90}, we show the test set $\mathcal{S}$ along with the computed intervals $[o_{5\%},o_{95\%}]$ of the proposed approach. Note that the intervals are wider when there are trend changes in the output. Furthermore, figure \ref{fig:max_lik} shows an example of the value of the maximum likelihood ratio $L_\gamma$ as a function of $\gamma$ (in this case for the data set of $200$ points and interval $[o_{5\%},o_{95\%}]$).

\section{Conclusions} \label{section6}
This work presents a new approach to obtain an interval predictor to be used in nonlinear systems.
The methodology relies on a parameterized family of dissimilarity functions, that are used to estimate the probability density function of the system output, conditioned to last inputs and outputs.  A family of empirical probability density functions, parameterized by means of two parameters, is proposed. It is shown that the  proposed family encompasses the multivariable normal probability density function as a particular case. The methodology allows us to provide probabilistic interval predictions of the output of the system.  For a particular choice of the tuning parameters, the conditional probability density function of the output attains a maximum at the output estimated by least-squares regression. This shows that the proposed method constitutes a generalization of classical estimation methods. A validation scheme is used to tune the two parameters on which the methodology relies ($c$ and $\gamma$). The method has been applied to generate interval predictions, which have been compared favourably with the ones obtained by means of quantile regression and set-membership methods.

\section*{Appendix A}
\textbf{Proof of Property 2}:
Denote $\lambda = [\lambda_1 \; \lambda_2 \; ... \; \lambda_N]\T$. We solve the optimization problem using a dual formulation where $\mu\in \R^{n+1}$ denotes the multipliers associated with the equality constraint
$$ z=\Sum{i=1}{N}\lambda_i z_i = Z\lambda,$$ and $\nu$ is the multiplier corresponding to the equality
 $$ 1 = \Sum{i=1}{N} \lambda_i = u\T \lambda.$$
  the Lagrange function is $$\mathcal{L}(\lambda,\mu,\nu) = \lambda\T\lambda + \mu\T(Z\lambda-z) +\nu(u\T\lambda-1).$$
  Denote $\lambda^*$, $\mu^*$ and $\nu^*$ the optimal values for the primal and dual variables.
From $\frac{\partial \mathcal{L}(\lambda^*,\mu^*,\nu^*)}{\partial \lambda}=0$ we obtain that the optimal vector $\lambda^*$ is given by
\begin{equation}\label{ec:lambda}\lambda^* = -\frac{1}{2}(Z\T\mu^*+u\nu^*).\end{equation}
Since $u\T \lambda^*=1$, $Zu=N\bar{z}$ and $u\T u=N$ we can premultiply both terms of last equality by $u\T$ to obtain
\begin{eqnarray*}
 1 &=& -\frac{1}{2}(u\T Z\T \mu^*+ N\nu^*)\\
 & = & -\frac{N}{2}(\bar{z}\T\mu^* + \nu^*).
\end{eqnarray*}
Therefore,
$$\nu^* =  -\frac{2}{N}-\bar{z}\T\mu^*.$$
Substituting the expression for $\nu^*$ in (\ref{ec:lambda}) yields,
\begin{eqnarray}
\lambda^* &=&  -\frac{1}{2}\left(Z\T\mu^*-u(\frac{2}{N}+\bar{z}\T\mu^*)\right) \nonumber\\
&=&\frac{u}{N} - \frac{1}{2}(Z\T- u\bar{z}\T )\mu^*.\label{ec:lambda2}
\end{eqnarray}
Premultiplying by $Z$ we obtain
\begin{equation}\label{ec:Zlambda} Z\lambda^* = \bar{z} - \frac{1}{2}(ZZ\T-N\bar{z}\bar{z}\T)\mu^*.\end{equation}
From the equality constraint $Z\lambda^*=z$ and (\ref{ec:Zlambda}) we have
$$\mu^* = -2(ZZ\T-N\bar{z}\bar{z}\T)^{-1}(z-\bar{z}).$$
Substituting $\mu^*$ in equation (\ref{ec:lambda2}) we infer
$$\lambda^* =  \frac{u}{N} + ( Z\T-u\bar{z}\T)(ZZ\T-N\bar{z}\bar{z}\T)^{-1}(z-\bar{z}).$$
Finally, taking into account that
$$u\T(Z\T-u\bar{z}\T  ) = (N\bar{z}\T-N\bar{z}\T)=0$$
we obtain
$$(Z\T -u\bar{z}\T )\T(Z\T -u\bar{z}\T) = ZZ\T-N\bar{z}\bar{z}\T.$$
From last equality and the expression for $\lambda^*$ we conclude
\begin{eqnarray*}
J_{0}(z,\mathcal{D}) &=& (\lambda^*)\T\lambda^* \\
& =& \frac{1}{N}+(z-\bar{z})\T(ZZ\T-N\bar{z}\bar{z}\T)^{-1}(z-\bar{z}). \;\quad \blacksquare
\end{eqnarray*}

\section*{Appendix B}

\textbf{Proof of Property 3}:
From equation (\ref{equ:optimal:y}) we have that the optimal value for the estimation is $\hat{y}_k = y^* = \Sum{i=1}{N} \lambda_i^* y_i$, where for the particular case $\gamma=0$, $\lambda_i^*$, $i=1,\ldots,N$, are the optimal values of the optimization problem
\begin{eqnarray*}
& \min\limits_{\lambda_1,\ldots,\lambda_N}  & \Sum{i=1}{N} \lambda_i^2  \\
& s.t. & x_k =\Sum{i=1}{N}\lambda_i x_i  \\
&& 1 = \Sum{i=1}{N}\lambda_i.
\end{eqnarray*}
Defining $$ R = \bmat{cccc} x_1 & x_2 & \ldots & x_N \\ 1 & 1 &\ldots & 1\emat,\;\; r_k = \bv x_k \\ 1 \ev,$$
we have that the equality constraints can be rewritten as
 \begin{equation}\label{equ:V:equality} R\lambda=r_k,
 \end{equation}
 where $\lambda=\bmat{cccc} \lambda_1 & \lambda_2 & \ldots & \lambda_N \emat\T$. From the Karush-Kuhn-Tucker optimality conditions we infer that the optimal solution is given by (see subsection 10.1.1 in \cite{Boyd04})
$$ \bmat{cc} \Id & R\T \\ R & 0 \emat \bv \lambda^* \\  \varphi^* \ev = \bv 0 \\ r_k\ev,
$$
where $\varphi^*$ corresponds to the optimal dual decision variables corresponding to the equality constraint (\ref{equ:V:equality}), (see \cite{Boyd04}). The previous equation can be rewritten as
\begin{eqnarray*}
 \lambda^* &=& - R\T \varphi^* \\
 R \lambda^* & = &r_k.
\end{eqnarray*}
From here we obtain $-RR\T \varphi^* =r_k$ which implies $\varphi^*=-(RR\T)^{-1} r_k$.
We finally obtain
$$ \lambda^* =R\T(RR\T)^{-1}r_k.$$
Therefore,
\begin{eqnarray*}
 \hat{y}_k &=& Y\T \lambda^* \\
 & = & Y\T  R\T (R R\T)^{-1} r_k\\
 & = & r_k\T (RR\T)^{-1}R Y,
 \end{eqnarray*}
where $Y=\bmat{cccc} y_1 & y_2 & \ldots & y_N\emat\T$. We notice that this corresponds to the least squares estimation obtained when we consider as regressors the vectors $\bmat{cc} x_j\T & 1 \emat\T$, $j=1,\ldots,N$ (see \cite{Ljung99}, \cite{Murphy:12}).
\hfill \QED

\bibliography{BibliografiaPercentil}

\end{document}